\documentclass[twocolumn,aps,prl,showpacs]{revtex4}

\usepackage{epsfig}
\usepackage{graphicx}
\usepackage{amsmath}
\usepackage{bm}

\begin{document}

\title {Nanoscale Impurity Structures on the Surface of $d_{x^2-y^2}$-wave Superconductors}
\author{Nikolaos A. Stavropoulos and Dirk K.~Morr}
\affiliation{Department of Physics, University of Illinois at
Chicago, Chicago, IL 60607}
\date{\today}
\begin{abstract}
We study the effects of nanoscale impurity structures on the local
electronic structure of $d_{x^2-y^2}$-wave superconductors. We show
that the interplay between the momentum dependence of the
superconducting gap, the geometry of the nanostructure and its
orientation gives rise to a series of interesting quantum effects.
Among these are the emergence of a zero bias conductance peak in the
superconductor's density of states and the suppression of impurity
states for certain nanostructures. The latter effect can be used to
screen impurity resonances in the superconducting state.

\end{abstract}


\maketitle

The study of impurities in the cuprate superconductors has attracted
significant interested over the last few years due to the
impurities' dramatic effects on the superconductor's local
electronic structure. In particular, fermionic resonance states
induced by single, isolated impurities have been well characterized
experimentally \cite{exp1,Hud01} and investigated theoretically
\cite{Sal96,theory1} (for a recent review, see
Ref.~\cite{Balatsky_RMP}). It was recently argued that quantum
interference effects involving two impurities yield new insight into
the nature of the electronic correlations in the cuprate
superconductors \cite{2imp}. At the same time, studies of nanoscale
impurity structures (so-called quantum corrals) in metals
\cite{Man00,theory,review} have led to the discovery of exciting
quantum effects, such as quantum imaging \cite{Man00} and similar
effects are also predicted to occur in $s$-wave superconductors
\cite{corral}. The question thus naturally arises whether
nanostructures in $d_{x^2-y^2}$-wave superconductors give rise to
novel quantum effects and provide new information on the cuprates'
complex electronic structure.

In this Letter, we address this question and study the effects of
nanoscale impurity structures on the local electronic structure of a
$d_{x^2-y^2}$-wave superconductor. We take the nanostructures to be
located on the superconductor's surface and to interact with the
superconductor's electronic degrees of freedom via a non-magnetic
scattering potential. We show that nanostructures induce fermionic
resonance states inside the superconducting gap, whose spatial form
is determined by the interplay between the momentum dependence of
the superconducting gap, the geometry of the nanostructure, and its
orientation with respect to the underlying lattice. This interplay
leads to a number of interesting quantum effects. First, a zero-bias
conductance peak (ZBCP) emerges in the superconductor's density of
states (DOS) whose dependence on the nanostructure's orientation
provides a new tool for identifying the symmetries of unconventional
superconductors. Second, for certain nanostructures, destructive
quantum interference leads to a complete suppression of impurity
resonances; an effect that can be used to spatially screen impurity
resonances in the superconducting state. Third, we demonstrate by
using more complex nanostructures, that it is possible to
``custom-design" the spatial form of impurity resonances. This
effect potentially provides a new probe for electronic correlations
of complex systems in general.

In order to study how a nanostructure affects the superconductor's
local electronic structure, we compute the superconductor's real
space Greens function within a generalized ${\hat T}$-matrix
scattering theory \cite{Morr04,Shiba68}. Introducing the spinor
$\Psi^\dagger_{{\bf r}}=\left(c^\dagger_{{\bf r},\uparrow},c_{{\bf
r},\downarrow} \right)$ the electronic Greens function,
$\hat{G}({\bf r}, {\bf r^\prime},\tau-\tau^\prime) =- \langle {\cal
T} \Psi_{{\bf r}}(\tau) \Psi^\dagger_{{\bf r^\prime}}(\tau^\prime)
\rangle$, in Matsubara frequency space is given by
\begin{eqnarray}
\hat{G}({\bf r},{\bf r'},\omega_n)&=&\hat{G}_0({\bf r},{\bf r'},\omega_n) \nonumber \\
& & \hspace{-2cm} +\sum_{i,j=1}^N \hat{G}_0({\bf r},{\bf
r}_i,\omega_n)\hat{T}({\bf r}_i,{\bf r}_j,\omega_n)\hat{G}_0({\bf
r}_j,{\bf r'},\omega_n) \ , \label{Ghat}
\end{eqnarray}
where the sum runs over the locations ${\bf r}_i \ (i=1,..,N)$ of
the $N$ impurities forming the nanostructure. The ${\hat T}$-matrix
is obtained from the Bethe-Salpeter equation
\begin{eqnarray}
\hat{T}({\bf r}_i,{\bf r}_j,\omega_n)&=& \hat{V}_i \delta_{i,j}  \nonumber \\
& & \hspace{-1cm} +\hat{V}_i \sum_{l=1}^N \hat{G}_0({\bf r}_i,{\bf
r}_l,\omega_n)\hat{T}({\bf r}_l,{\bf r}_j,\omega_n) \ ,
\label{Tmatrix}
\end{eqnarray}
where ${\hat V}_i=U_i \sigma_3$, $U_i$ is the non-magnetic
scattering potential of the impurity at site ${\bf r}_i$, and $
\sigma_j$ are the Pauli-matrices. We consider identical impurities
and take for definiteness $U_i=1$eV, however, the results discussed
below remain qualitatively unchanged over a wide range of scattering
strength. The electronic Greens function of the unperturbed (clean)
superconductor in momentum space is given by
\begin{equation}
\hat{G}^{-1}_0({\bf k},i\omega_n)= i\omega_n \sigma_0 -
\varepsilon_{\bf k} \sigma_3 + \Delta_{\bf k} \sigma_1 \ ,
\label{G0}
\end{equation}
where $\Delta_{\bf k}=\Delta_0\left(\cos k_x - \cos k_y \right)/2$
is the superconducting $d_{x^2-y^2}$-wave gap with $\Delta_0=25$
meV, and
\begin{equation}
\varepsilon_{\bf k}=-2t \left( \cos k_x + \cos k_y \right) - 4 t'
\cos k_x \cos k_y - \mu \label{Disp}
\end{equation}
is the host system's normal state tight binding dispersion with
$t=300$ meV, $t'/t=-0.4$, and $\mu/t=-1.18$, representative of the
cuprate superconductors \cite{ARPES}. The local DOS, $N({\bf
r},\omega)$, is obtained numerically from
Eqs.(\ref{Ghat})-(\ref{Disp}) with $N({\bf r},\omega)=-2{\rm Im}\,
\hat{G}_{11}({\bf r},\omega+i\delta)/ \pi$ and $\delta=0.2$ meV.

\begin{figure}[h]
\epsfig{file=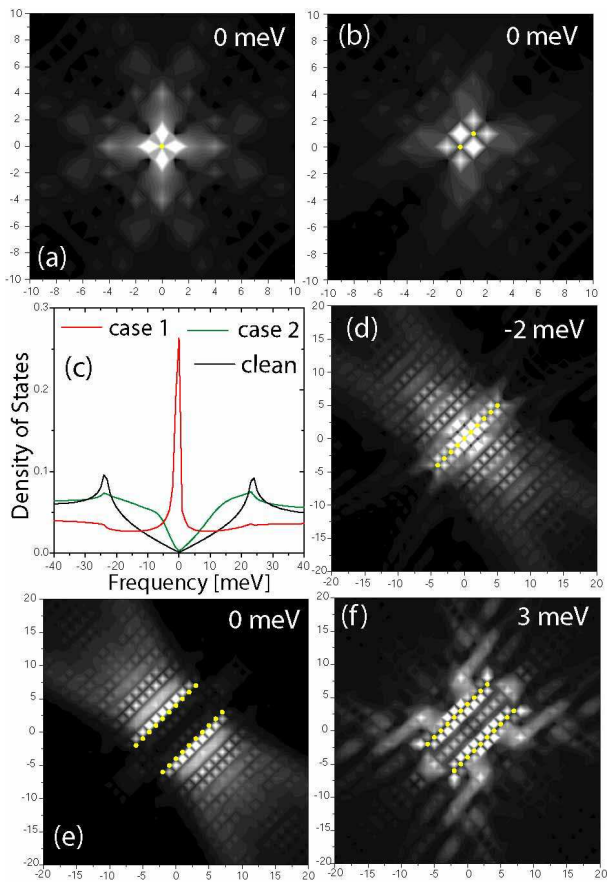,width=8.0cm} \caption{Intensity plot of the
DOS for (a) a single impurity located at ${\bf r}_1=(0,0)$, (b) two
impurities located at ${\bf r}_1$ and ${\bf r}_2=(1,1)$, (d) $N=10$
impurities aligned along the (110)-direction, (e),(f) two lines with
$N=10$ impurities each separated by $d=4\sqrt{2}$ (the lattice
constant is set to $a_0=1$). (c) DOS as a function of frequency at
$(0,1)$ (see text). Non-magnetic impurities are represented by
filled yellow circles. }\label{Evolution}
\end{figure}

In order to better understand the dependence of impurity states on
the geometry and orientation of a nanostructure, it is necessary to
first study how the spatial form of impurity resonances evolves as
the size and complexity of a nanostructures increases. As a
reference point, we plot in Fig.~\ref{Evolution}(a) the spatial DOS
pattern of the resonance state at $\omega=0$ meV that is induced by
a single impurity located at ${\bf r}_1=(0,0)$ [light (dark) color
indicates a large (small) DOS].
%
%
When a second impurity is added, the mere existence of an impurity
state depends on the orientation of this impurity dimer relative to
the underlying lattice. If the two impurities are aligned along the
(110)-direction (case I), with the second impurity at ${\bf
r}_2=(1,1)$, a resonance state exists at $\omega=0$ with the spatial
DOS structure shown in Fig.~\ref{Evolution}(b). In contrast, if the
second impurity is placed at ${\bf r}'_2=(0,1)$ (case II) and the
dimer is aligned along the (100)-direction, the DOS exhibits only
Friedel-like oscillations, but no impurity resonance. This striking
difference is particularly apparent when one plots the DOS at
$(0,1)$, as shown in Fig.~\ref{Evolution}(c). This dependence of the
resonance on the dimer's orientation is similar to that of the zero
bias conductance peak (ZBCP) observed near one-dimensional surface
edges in the cuprate superconductors \cite{ZBCP}. Only if the
electrons that are specularly scattered along the impurity dimer
experience a sign change in the superconducting gap, a ZBCP-like
impurity resonance emerges, such as the one shown in
Fig.~\ref{Evolution}(b) for case I. This orientational dependence is
an important feature of nanostructures, as further discussed below.

A characteristic signature of the ZBCP-like state is that it extends
spatially perpendicular to the impurity line, as shown in
Fig.~\ref{Evolution}(d). Note, however, that the ZBCP-like state is
shifted away from zero energy, and located at $\omega=\pm 2$ meV.
This shift arises from the hybridization of the ZBCP-like states on
both sides of the impurity line, and the resulting formation of
bonding and antibonding resonances. This hybridization is mediated
by the next-nearest neighbor hopping term (the $t^\prime$-term),
which permits the exchange of electrons between the two sides of the
impurity line without a scattering process. When the hybridization
of the ZBCP-like states is suppressed, as for example, in the
nanostructure consisting of two parallel lines shown in
Fig.~\ref{Evolution}(e), the ZBCP-like state is again located at
zero-energy. Here, the ZBCP-like states are spatially separated and
the absence of an impurity state between the two lines prevents the
coupling of the states and thus their hybridization. A similar
effect is also observed for other nanostructures (see below). In
addition to a ZBCP-like state, the nanostructure also induces
impurity states with different spatial patterns, such as the
``frog-like" resonance shown in Fig.~\ref{Evolution}(f). Quite
interestingly, the global spatial pattern of this resonance extends
along the (100)-direction, but consists of lines with increased DOS
along the (110)-direction. Finally, note that the intensity of the
ZBCP-like states decreases algebraically $\sim 1/r^\alpha$ along the
(110)-direction with distance from the nanostructure. For the single
impurity resonance in Fig.~\ref{Evolution}(a), one has $\alpha=2$
\cite{Sal96}, while for nanostructures and distances smaller than
their lateral size, we find in general $\alpha < 2$. For example,
$\alpha \approx 1.18$ for the nanostructure shown in
Fig.~\ref{Evolution}(e). This decrease of $\alpha$ with increasing
length of the impurity line is expected since for a truly
one-dimensional ZBCP-like state at zero energy, one has $\alpha=0$.

\begin{figure}[h]
\epsfig{file=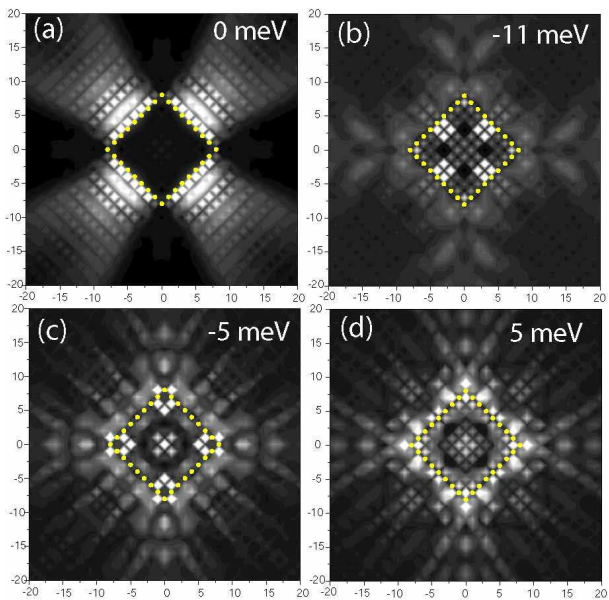,width=8.0cm} \caption{ Intensity plot of the
DOS for a square with $N=32$ impurities and sides of length
$a=8\sqrt{2}$ which are parallel to the (110) and symmetry related
directions.} \label{DOS_square}
\end{figure}

To further explore the interplay between geometry and orientation,
we next consider nanostructures in the form of a square, such as the
one shown in Fig.~\ref{DOS_square} whose sides are parallel to the
(110)-direction.
%
%
This nanostructure induces a ZBCP-like resonance at $\omega=0$ meV
[Fig.~\ref{DOS_square}(a)], which extends in all four equivalent
(110)-directions. The ZBCP-like resonances associated with each side
are spatially separated and thus do not hybridize. Impurity states
at non-zero frequencies exhibit different characteristic spatial
patterns. The impurity resonance shown in Fig.~\ref{DOS_square}(b)
extends along the (100)-direction outside the square with a
``V-like" fine structure consisting of branches that run along the
(110)-direction. In contrast, the impurity state shown in
Fig.~\ref{DOS_square}(c) and (d) is ``star-like" and extends almost
radially outwards. Note that the ``global" (i.e., larger length
scale) spatial structure of the DOS at $\omega=\pm 5$ meV
[Figs.~\ref{DOS_square}(c) and (d)] is identical, and thus
particle-hole symmetric. This symmetry also exists for other
impurity resonances, such as the ones at $\omega= \pm 11$ meV. Only
in the local, i.e., small length scale structure of the impurity
resonance can we identify differences in the spatial DOS pattern. Of
particular interest is the (local) $45^\circ$ rotation of the DOS
intensity between particle-like and hole-like energies in the
vicinity of the nanostructure, which is similar to that observed
near single impurities in the cuprate superconductors \cite{Hud01}.
We find that the dichotomy of global particle-hole symmetry and
local particle-hole asymmetry of impurity resonances is also
exhibited by other nanostructure geometries.

\begin{figure}[h]
\epsfig{file=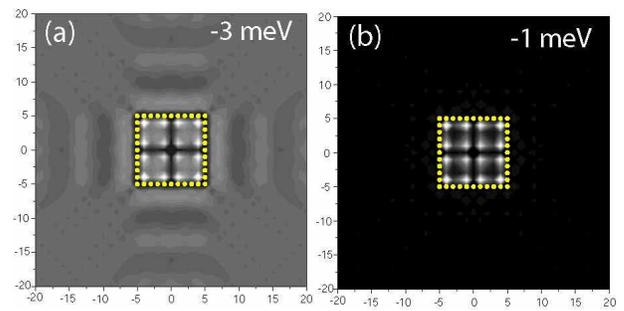,width=8.0cm} \caption{ Intensity plot of the
DOS for a square with $N=40$ impurities in (a) the normal state, and
(b) the superconducting state.}\label{100square}
\end{figure}

When the square is rotated by $45^\circ$ such that its sides are
parallel to the (100) direction [see Fig.~\ref{100square}],
specularly reflected electrons do not experience a sign change in
the superconducting pairing potential and hence no impurity
resonances are formed. This orientational dependence of the impurity
resonances demonstrates that nanostructures are a new tool to
identify the symmetry of unconventional superconductors in general.
%
%
We find, however, that if an eigenmode of the nanostructure exists
in the normal state at energies $|\omega| < \Delta_0$, the same
eigenmode can also be excited in the superconducting state. Consider
for example the normal state eigenmode at $\omega=-3$ meV shown in
Fig.~\ref{100square}(a). In the superconducting state, the same
eigenmode is excited at $\omega=-1$ meV, as shown in
Fig.~\ref{100square}(b). Note that this eigenmode in the
superconducting state does {\it not} arise from pairbreaking
effects, such as for example, the impurity {\it pairbreaking}
resonances shown in Figs.~\ref{Evolution}(a),(b). It is therefore
important to distinguish between impurity states that arise from
eigenmodes of the nanostructure and those that are pairbreaking
resonances. The spatial structure of impurity states is often
determined by an interplay of eigenmodes and pairbreaking
resonances, with eigenmodes (pairbreaking resonances) determining
the spatial DOS pattern inside (outside of) the nanostructure.

%
%
\begin{figure}[h]
\epsfig{file=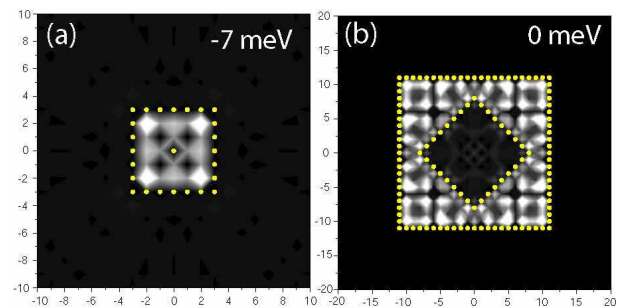,width=8.0cm} \caption{Intensity plot of the
DOS for (a) a screened single impurity, and (b) a screened impurity
square (see text). }\label{screening}
\end{figure}
The fact that impurity states are absent for nanostructure squares
whose sides are parallel to the (100)-direction can be used to
``screen" impurity states in the superconducting state. For example,
when a single impurity is enclosed by a square, as shown
Fig.~\ref{screening}(a), the impurity state is completely confined
to the interior of the square. This screening process considerably
reduces the spatial extent of the impurity resonance, as follows
from a comparison of Figs.~\ref{Evolution}(a) and
\ref{screening}(a). Even the more complex ZBCP-like state of
Fig.~\ref{DOS_square}(a) can be screened when enclosed by a second
square, as shown in Fig.~\ref{screening}(b).

\begin{figure}[h]
\epsfig{file=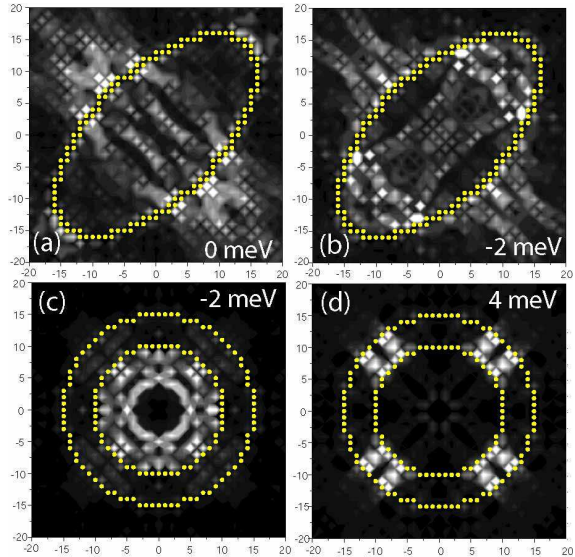,width=7.5cm} \caption{Intensity plot of the
DOS for (a),(b) an elliptical quantum corral with $N=84$ impurities,
whose axes of length $a=20$ and $b=10$ are aligned along the
(110)-direction, and (c),(d) two congruent circles with radii
$R_1=10$ and $R_2=15$ and $N=141$ impurities. }
\label{CircleEllipse}
\end{figure}

Ellipses, whose major axis are aligned along the (110)-direction,
also possess ZBCP-like impurity states, as shown in
Fig.~\ref{CircleEllipse}(a) and (b). In particular, at $\omega=0$
meV, the impurity state permeates the entire ellipse along the minor
axis of the ellipse. In contrast, at $\omega=-2$ meV, the impurity
resonance outside the ellipse extends parallel to its minor axis,
while in the ellipse's interior, is extends along its major axis.
These ZBCP-like resonances are absent when the axes of the ellipse
are parallel to the (100)-direction.
%
%
More complex nanostructures provide the possibility to create
impurity resonances that are confined to distinctively different
parts of the nanostructure. For example, the nanostructure
consisting of two congruent circles shown in
Figs.~\ref{CircleEllipse}(c) and (d), possesses an impurity
resonance at $\omega=-2$ meV [Figs.~\ref{CircleEllipse}(c)] that is
located predominantly inside the inner circle, while the one at
$\omega=4$ meV [Fig.~\ref{CircleEllipse}(d)] which is reminiscent of
the ZBCP, is confined to the area between the two circles. This
result exemplifies the possibility to ``custom-design" the spatial
structure of impurity resonances, with the potential of gaining
further insight into the complex electronic structure of the
cuprates.

Since non-magnetic impurities are pair-breaking in a
$d_{x^2-y^2}$-wave superconductor, we expect that nanostructures
also lead to a spatial variation of the superconducting order
parameter. The effects of this spatial variation are beyond the
scope of the present work, but will be addressed in future work
within the self-consistent Bogoliubov de Gennes formalism. Our study
of nanostructures in $s$-wave superconductors, however, has shown
that such a spatial variation leads only to small quantitative
changes from the results of the ${\hat T}$-matrix theory
\cite{Morr05}. We therefore expect that our above results remain
qualitatively unchanged when the spatial variation of the
superconducting order parameter is taken into account.

In summary, we have studied the effects of nanostructures on the
local electronic structure of a $d_{x^2-y^2}$-wave superconductor.
We show that the interplay between the geometry of the
nanostructure, its orientation and the momentum dependence of the
superconducting gap gives rise to a series of interesting quantum
effects, such as the emergence of ZBCP-like states and the screening
of impurity resonances. These effects demonstrate that
nanostructures are a new tool for identifying the symmetry of
unconventional superconductors and for probing the electronic
correlations of complex systems.

We would like to thank J.C. Davis for stimulating discussions.
D.K.M. acknowledges financial support from the Alexander von
Humboldt foundation.

\end{document}